\documentstyle[11pt,aaspp4]{article}
\received{16 September 1998}
\accepted{14 December 1998}
\lefthead{Yu et al.}
\righthead{Kilohertz QPO Frequency and Flux Decrease in Aql X-1 and 
Effect of Soft X-ray Spectral Components
}
\begin{document}
\title{Kilohertz QPO Frequency and Flux Decrease in AQL X-1 and Effect 
of Soft X-ray Spectral Components}
\author{W. Yu\altaffilmark{1}, 
T. P. Li\altaffilmark{1},
W. Zhang\altaffilmark{2},
S. N. Zhang\altaffilmark{3,4}}
\altaffiltext{1}{LCRHEA, Institute of High Energy Physics, Beijing 100039, 
China }
\altaffiltext{2}{ Laboratory for High Energy Astrophysics, Goddard Space 
Flight Center, Greenbelt, MD 20771 } 
\altaffiltext{3}{ University of Alabama in Huntsville, Department of
Physics, Huntsville, AL 35899 } 
\altaffiltext{4}{ NASA/Marshall Space Flight Center, ES-84, Huntsville, 
AL 35812 } 

\begin{abstract} 

We report on an RXTE/PCA observation of Aql X-1 during its outburst in 
March 1997 in which, immediately following a Type-I burst, the 
broad-band 2-10 keV flux decreased by about 10\% and the kilohertz 
QPO frequency decreased from 813$\pm$3 Hz to 776$\pm$4 Hz. This change 
in kHz QPO frequency is much larger than expected from a simple 
extrapolation of a frequency-flux correlation established using data 
before the burst. Meanwhile a very low frequency noise (VLFN) 
component in the broad-band FFT power spectra with a fractional 
root-mean-square (rms) amplitude of 1.2\% before the burst ceased to 
exist after the burst. All these changes were accompanied by a change 
in the energy spectral shape. If we characterize the energy spectra 
with a model composed of two blackbody (BB) components and a power law 
component, almost all the decrease in flux was in the two BB components. We 
attribute the two BB components to the contributions from a region 
very near the neutron star or even the neutron star itself and from the 
accretion disk, respectively.

\end{abstract}

\keywords{X-ray: stars - stars: individual (Aquila X-1) - stars: neutron}

\section{Introduction}

Kilohertz QPOs have been observed in about 20 low-mass X-ray binaries 
by the {\it Rossi X-ray Timing Explorer} (RXTE) 
since its launch at the end of 1995 (\cite{klis97}). Although the detail 
production mechanism of these QPOs is not fully understood, there is little
doubt that they correspond to the  dynamical time scale near the
neutron star surface, and as such they enable us to probe strong
gravity effects and the equation of state of neutron stars
(\cite{kaaret97}; \cite{zhangw97}; \cite{klis97}; \cite{miller97};
\cite{lamb98}).

The kHz QPO frequency has been found to correlate with at least two
quantities: source count rate or flux and energy spectral shape or
hardness ratios (\cite{klis97} and references therein; \cite{kaaret98}; 
\cite{zhangw98b}; \cite{mendez99}). Soft X-ray transients like 4U 1608-52 and
Aql X-1, with their large swings in both count rates and energy spectral 
shapes, are ideal for the study of these kinds of correlations. 
4U 1608-52 and Aql X-1 have both been observed with RXTE/PCA during their outburst
decay. A similar correlation between the QPO frequency and the X-ray
flux has been observed in different flux ranges (\cite{yu97};
\cite{mendez98}; \cite{zhangw98}).  In each flux range, the QPO
frequency appears correlated with the X-ray flux. Spectral studies
suggest that there is a significant correlation between the QPO
frequency and the spectral shape even among data from different flux
ranges in 4U 1608-52 (\cite{kaaret98}). A correlation between the QPO
frequency and the spectral shape has also been found in other QPO
sources. For example, a correlation between the QPO frequency and the
flux of the blackbody component was observed in 4U 0614+091
(\cite{ford97}).

The QPO frequency is usually regarded as the Keplerian frequency at the
inner disk radius or its beat frequency with the neutron star spin in
the beat-frequency model (\cite{alpar85}; \cite{lamb85};
\cite{strohmayer96}; \cite{zhangsn98}; \cite{miller96}).  The higher
the mass accretion rate, the higher should be the QPO frequency.  This
will lead to a correlation between the QPO frequency and the flux.
However, when a transition of the accretion state happens, same mass
accretion rate may correspond to different inner disk radii and
different fluxes before and after the transition. A spectral variation
is probably associated with a variation of the accretion state, and
might lead to a new flux range or frequency range. Comparison of the
spectral components before and after the spectral variation and study
of the correlation between QPO frequency and the spectral shape will
help to reveal how the kHz QPOs are generated.

In this letter, we present our study with the RXTE/PCA data before and 
after an X-ray burst in Aql X-1. 

\section{Observations and Data Analysis}

An X-ray burst with a peak PCA count rate above 40,000 cps was
observed by RXTE/PCA on March 1, 1997, when the daily-averged ASM count 
rate is $\sim$ 4 cps. Nearly coherent oscillation
at about 549 Hz was detected in the X-ray burst. A $\sim$10\% flux
decrease and a corresponding QPO frequency decrease were observed
after the X-ray burst (\cite{zhangw98}). The count rate decrease 
is $\sim$ 64 cps in the entire PCA band. The data used
for timing analysis in the kHz frequency range is taken in the event mode {\it
E\_125us\_64M\_0\_1s}. The energy range 2-10 keV was selected, the same
as those used in Zhang et al. (1998a). The {\it Standard 1} data is used
to study the frequency range below 0.5 Hz.  The {\it Standard 2}
data is used in the spectral analysis. All the errors quoted in the
following are 1 $\sigma$ errors. The low energy response of the
RXTE/PCA detectors is not sufficient in constraining the absorption
column density $N_{H}$. We have fixed it at 3.0$\times$${10}^{21}$
{cm}$^{-2}$ in our spectral fittings,  according to previous studies
(\cite{czerny86}; \cite{verbunt94}; \cite{zhangsn98}).

\subsection{Timing Analysis} 
We first calculate the dynamical power spectra around
the X-ray burst. Then a model composed of a constant and a Lorentzian
function is fit to the 200-2000 Hz power spectra to determine the QPO
frequency. In Fig. 1, we show the light curve around the X-ray burst and
how QPO frequency evolved. The QPO frequency decreased from about
813$\pm$3 Hz before the burst to 776$\pm$4 Hz after the burst.  In Fig.
2, the QPO frequency vs. PCA count rate (CR) in the energy range 2-10
keV is shown for the observation of two consecutive RXTE orbits on
March 1. The data points on the lower-left corner are taken 
after the X-ray burst, and those points on the right side
branch are taken before the X-ray burst. We apply a linear model
to fit the right side branch and get
$f_{qpo}=-(219.4\pm32.0)+(1.954\pm0.002)\times CR$.  The average count
rate after the X-ray burst in 2-10 keV band is $\sim$ 478 cps, the
inferred QPO frequency from the above QPO frequency vs. count rate
relation at this count rate is 714$\pm$30 Hz. Thus the inferred QPO
frequency decrease is about 99$\pm$30 Hz, more than twice of the
observed frequency decrease of 37$\pm$5 Hz.

In order to study the source variability at frequencies lower than 
0.5 Hz, we 
divide the light curve into 256 s segments, then calculate the power 
spectrum of each segment. We average 16 power spectra 
before the X-ray burst and 4 power spectra after the burst. In Fig. 3, 
we show the average Leahy normalized power spectra in the frequency range 
0.002-0.5 Hz before and after the X-ray burst. The power spectrum before 
the X-ray burst shows a very low frequency noise (VLFN) component of a 
power law form (\cite{HK89}). A model composed of a white noise level 
of 2 and a power law component is fit to the average power spectra. 
We obtain a power law index $\alpha$=$1.5\pm0.2$ and $A$=$0.0024\pm0.0015$. 
The fractional root-mean-square ({\it rms}) variability of the VLFN is 
about 1.2$\pm$0.4\%. The power spectrum after the X-ray burst is consistent 
with a white noise. 

\subsection{Spectral Analysis}
The energy spectrum of Aql X-1 has changed little during the outburst 
before March 1 (\cite{zhangw98}) and was of a blackbody type. In 
subsequent observations, the energy spectra gradually changed from a 
blackbody type to a blackbody plus a power-law type (see \cite{zhangsn98}). 
The spectra before and after the X-ray burst are shown in Fig. 4. 
In Fig. 4(a), we plot the ratio between the spectra 
(2-10 keV) before and after the X-ray burst, namely spectrum $N_{1}(E)$ 
and spectrum $N_{2}(E)$, respectively. The ratio as a function of energy 
is not a constant. The flux decrease is more severe above 5 keV. 
Thus a variation of the spectral shape was associated with the flux 
decrease. 

Spectral models of a single blackbody (BB), a single power law (PL) and 
their combination (BB+PL) can not yield an acceptable fit to 
both spectra $N_{1}(E)$ and $N_{2}(E)$ in the energy range 3-20 keV. 
The model composed of two BB components and a PL component gives an 
acceptable fit. The inclusion of two BB components, one of which may be 
a multi-color disk (MCD) component, has been applied 
previously to the soft X-ray transient 4U 1608-52 
during an outburst (\cite{mitsuda84}) and Aql X-1 during an outburst rise 
(\cite{cui98}). In Table 1, we list the parameters 
of the spectral fit to both spectra with 2BB+PL and BB+MCD+PL models. 
The emission area of the $\sim$ 0.26 keV BB component in the 2BB+PL model 
fit is much larger than the surface area of a neutron star. This suggests that 
it may be a disk component and the fit with BB+MCD+PL model may be reasonable. 
Assuming the disk inclination angle as zero, we find that the apparent inner 
disk radius is in the 1 $\sigma$ range 200-470 km. 
In both spectra, the PL components 
only contribute less than 1/20 of the total flux in the energy range 2-10 
keV. The difference between the average PCA count rate ($>$ 10 keV) before 
and after the X-ray burst is 0.64$\pm$0.38 cps. This indicates that the 
PCA flux variation mainly comes from the spectral components in the soft 
X-ray range below 10 keV. 

The two BB temperatures are stable during the flux decrease as shown in Table 1. 
Thus it is reasonable to study the flux decrease by subtracting the 
spectrum $N_{2}(E)$ from the spectrum $N_{1}(E)$. In Fig. 4(b), we plot the 
subtracted spectrum ($N_{1}(E)$-$N_{2}(E)$) in the energy range 2-10 keV. 
A model composed of two BB components yields an acceptable 
fit to this spectrum, which is also plotted. In Table 2, we list the best-fit 
parameters and the fluxes of the two components. The 2-10 keV energy flux 
of the residual spectrum is about (7$\pm$2)\% of that of the total 
before the X-ray burst. The subtraction of the PL components in 
Table 1 mainly affects the 0.28 keV component shown in Fig. 4(b), 
i.e. a flux of 0.7$\pm$2.5 photons cm$^{-2}$ s$^{-1}$ in 2-4 keV. 

\section{Discussion and Conclusion}
We have reported that there was a simultaneous decrease of the X-ray
flux, the QPO frequency, and the VLFN component in Aql X-1 immediately
following a Type-I X-ray burst.  The decrease lasted at least 400 s
until the observation was stopped. The flux decrease was mainly due to
a decrease of the spectral components in the energy range below 10 keV.
As a type I X-ray burst only occurs when the condition for the
thermonuclear instability is met, the X-ray burst and the flux decrease 
may be causally related.

The X-ray flux derived from the spectral fit in Table 1 has an
uncertainty as large as 20\%, which is insufficient to determine the
decrease of the X-ray luminosity. However, the spectral fit shown in
Table 2 yields more confined parameters, suggesting that there is a
(7$\pm$2)\% decrease of the X-ray luminosity in 2-10 keV.  The index of
the PL component decreases together with the decrease of the X-ray
luminosity.  This behavior is similar to that found in 4U 1608-52 and
4U 0614+091 (\cite{kaaret98}).

The spectral parameters of the two BB components in Table 1 seems 
consistent with the trend shown in Fig.2 of Cui et al. 1998, i.e. 
the lower the ASM count rate of Aql X-1, the larger the inner 
disk radius and the lower the BB temperatures are. 
The $\sim$ 1.06 keV BB is probably 
related to a part of the neutron star surface. 
The decrease of the VLFN after the burst supports the idea. 
The VLFN in LMXBs 
is an indication of the time-dependent fusion reactions 
(fires) on the neutron star surface (\cite{bildsten93}), which 
corresponds to a mass accretion rate larger than that 
of the bursting stage. The disappearance of the VLFN 
in Aql X-1 then not only suggests that at least a few percent 
of the observed X-ray flux was from the fusion reaction on the 
neutron star surface before the X-ray burst, but also indicates 
that after the X-ray burst, the time-dependent fusion reactions 
were probably stopped. This may indicate that the X-ray burst had 
consumed almost all the nuclear fuel on the neutron star, and 
the stop of the time-dependent fusion reaction contributes to a significant 
part of the 7\% X-ray flux decrease. 

The inner disk radii before and after the X-ray burst, derived from Table 1, 
are about 300 km and 360 km, respectively. For an accreting neutron star 
of 2.0 ${M}_{\odot}$ and taking $cos(\theta)$=1, the upper limits on 
the Keplerian frequencies corresponding to the two radii are 15 and 12 Hz. 
The spectral hardening would lead to an even larger inner disk radius
thus an even lower frequency (\cite{shimura95}). On the other hand, 
the QPO frequency around 800 Hz is probably the
lower of the twin peaks, which suggests that the Keplerian frequency at
the inner edge of the disk should be around 1075 Hz (\cite{zhangw98}). 
Thus the derived inner disk radii are inconsistent with the beat-frequency 
model. The difference
between the observed QPO frequency and those inferred from the spectral
fit probably comes from a lack of detailed knowledge of the spectral
components in neutron star X-ray binaries, and the lack of sensitivity
below 2 keV of the PCA to determine the parameters of a 0.26 keV
blackbody spectral component.  Conversely, it is also possible that the
identification of the kHz QPOs as the Keplerian frequency at the inner
edge of the accretion disk or its beat frequency against the neutron
star spin is incorrect.

The observed QPO frequency decrease is 37$\pm$5 Hz, about 1/3 of that 
inferred from the QPO frequency vs. count rate relation. 
In principle, the decrease of the disk BB flux should originate from a 
decrease of the mass accretion rate. This would introduce a decrease 
of the QPO frequency. 
Taking the spectral parameters in Table 2 and using the PCA instrumental 
information, we estimate that the 2-10 keV PCA count rate of the disk 
BB is 23$\pm$8 cps, and that of the neutron star BB component is 
47$\pm$6 cps. So the QPO frequency decrease is consistent 
with the 23 cps count rate decrease of the disk BB component, which is 
about 1/3 of the total decrease of $\sim$ 70 cps. Because the PCA 
effective area is not a constant with photon energy, 
and the two BB components have quite different BB temperatures, 
the incident BB photon fluxes (Table 2) can not replace the above 
count rate estimates when we study the frequency decrease inferred from 
the PCA count rate.   
  
The central radiation force may also affect the QPO frequency. A 
decrease of the
central BB emission from near the neutron star surface would probably
introduce an increase of the QPO frequency. Thus two mechanisms would
account for the correlation between the QPO frequency and the count
rate in Fig. 2. One is that it is the disk BB flux instead of the total
flux that is correlated with the QPO frequency. The QPO frequency 
variations between the data points on the right side branch in
Fig. 2 originate from the disk BB flux variations. Thus the data points
on the lower left corner may join in the data points of the right side
branch in a plot of the QPO frequency vs. the disk BB flux. The other
is that a decrease of the central BB radiation force after the X-ray
burst would increase the QPO frequency, then the QPO frequency did not
follow the correlation relation when the central BB radiation force is
nearly constant. Our study indicates that a comprehensive investigation
on the correlation between QPO frequency and each spectral component
is needed, especially when a flux decrease or spectral transition
occurs.

\begin{acknowledgments}
We thank an anonymous referee for helpful comments and suggestions, which 
certainly improved this article. 
This work was supported by the National Natural 
Science Foundation of China under grant 19673010 and 19733003. 
\end{acknowledgments}

%\end{document}  

\newpage

\begin{figure*}
\plotone{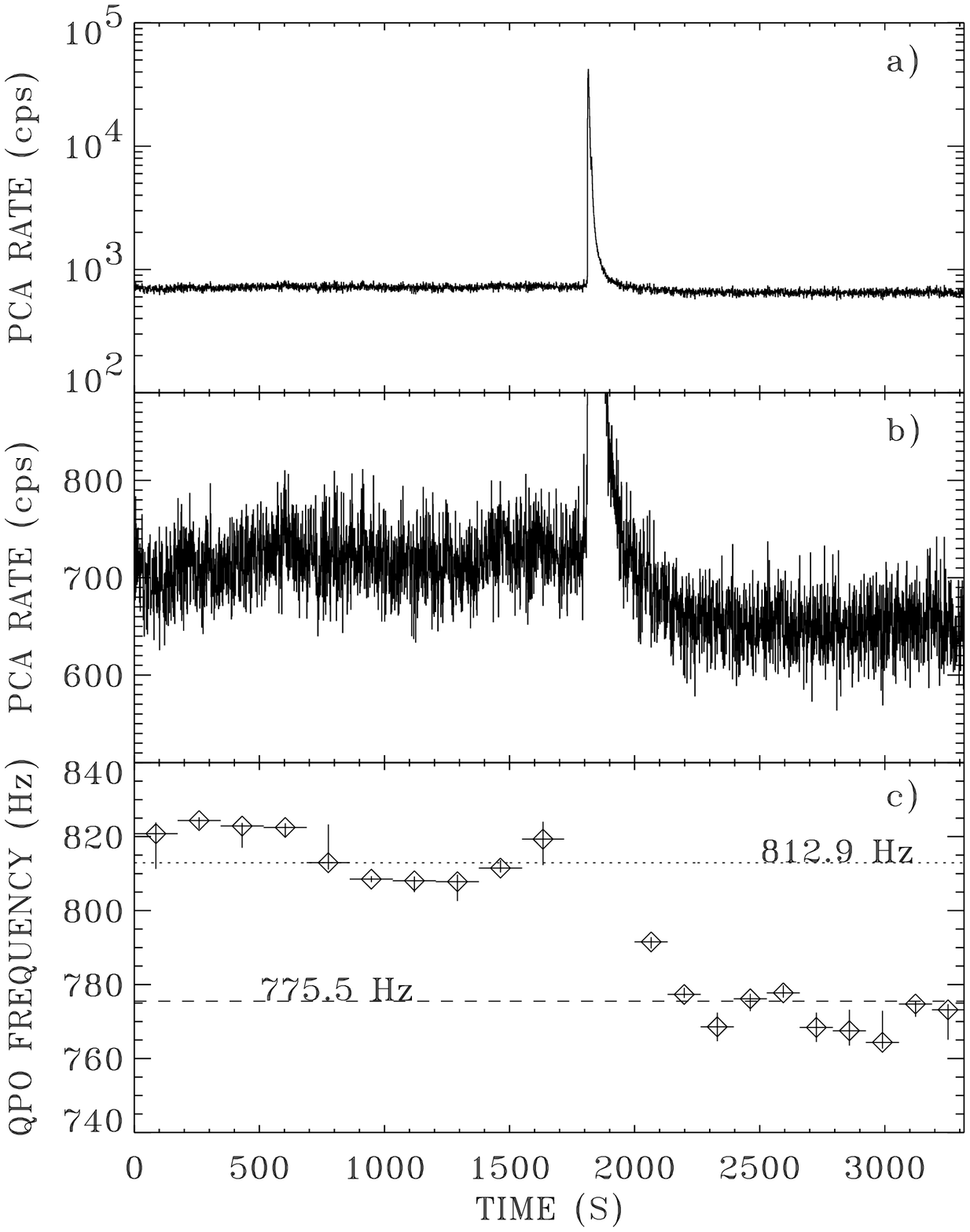}
\caption{Light curve and the kilohertz QPO evolution around the X-ray burst. 
a) The light curve 
in the entire PCA energy band ($\sim$ 2-60 keV) with background-subtraction. 
b) The flux decrease shown in detail; 
c) The QPO frequency evolution around the burst. }
\end{figure*}

\begin{figure*}
\plotone{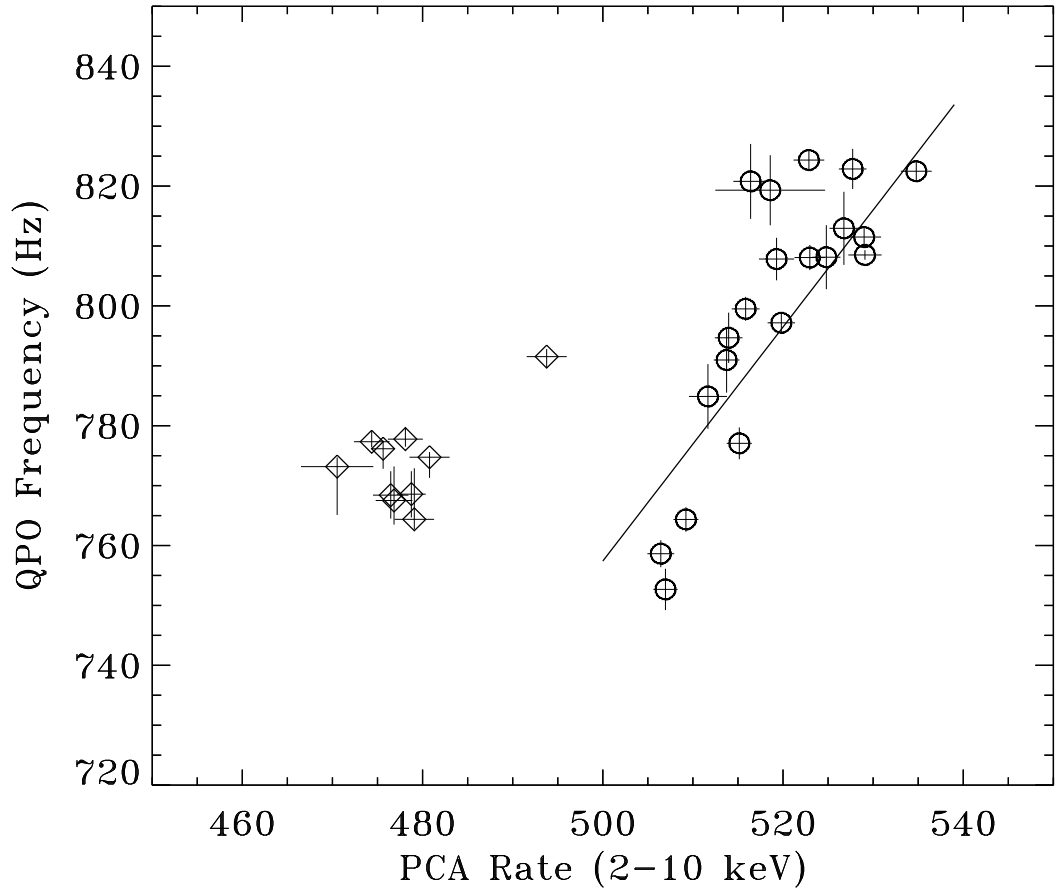}
\caption{Kilohertz QPO frequency vs. PCA count rate (2-10 keV) relation 
observed on March 1. The data points on the left (diamonds) are taken 
after the X-ray burst, and those on the right (circles) are taken before 
the burst.}
\end{figure*}

\begin{figure*}
\plotone{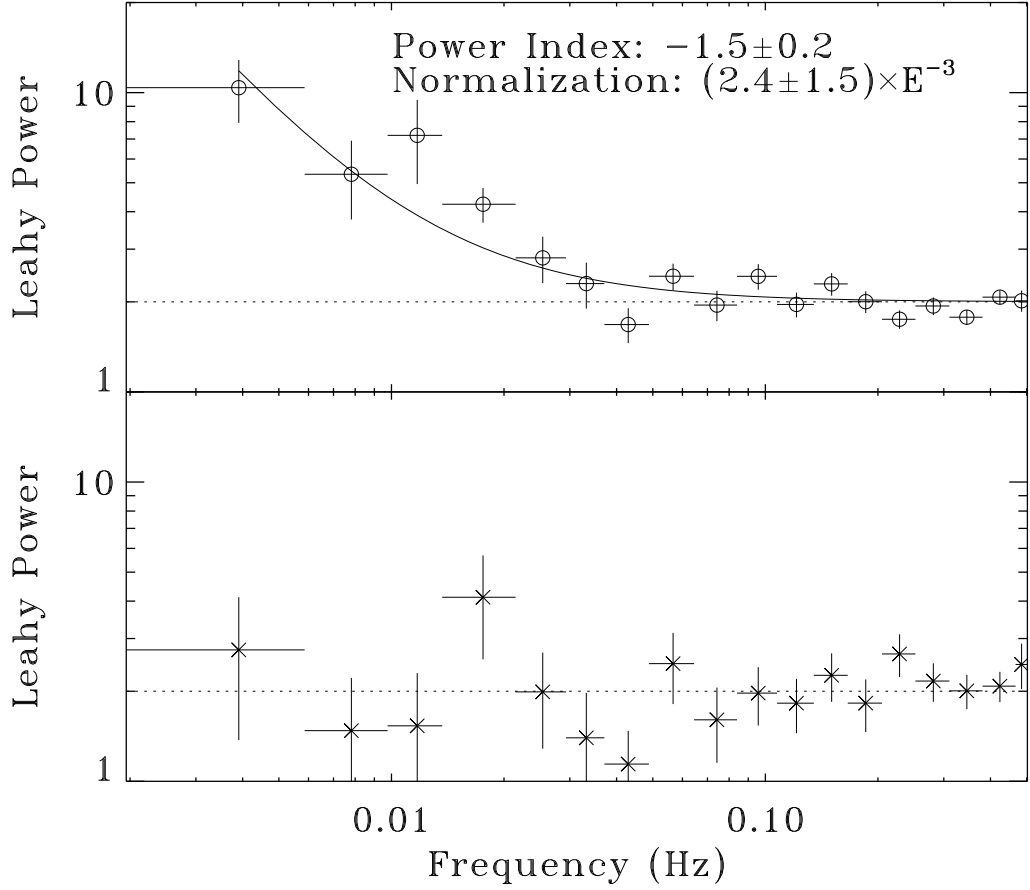}
\caption{The power spectra (0.002-0.5 Hz) for the 
entire PCA band before and after the 
X-ray burst. The upper panel and the lower panel correspond to 
data before and after the X-ray burst, respectively. The power spectrum 
in the upper panel shows a VLFN component, while the power spectrum of the 
lower panel agrees with a white noise. The best-fit model composed of 
a constant white noise level of 2 and a power law VLFN is shown in the 
upper panel.}
\end{figure*}

\begin{figure*}
\plotone{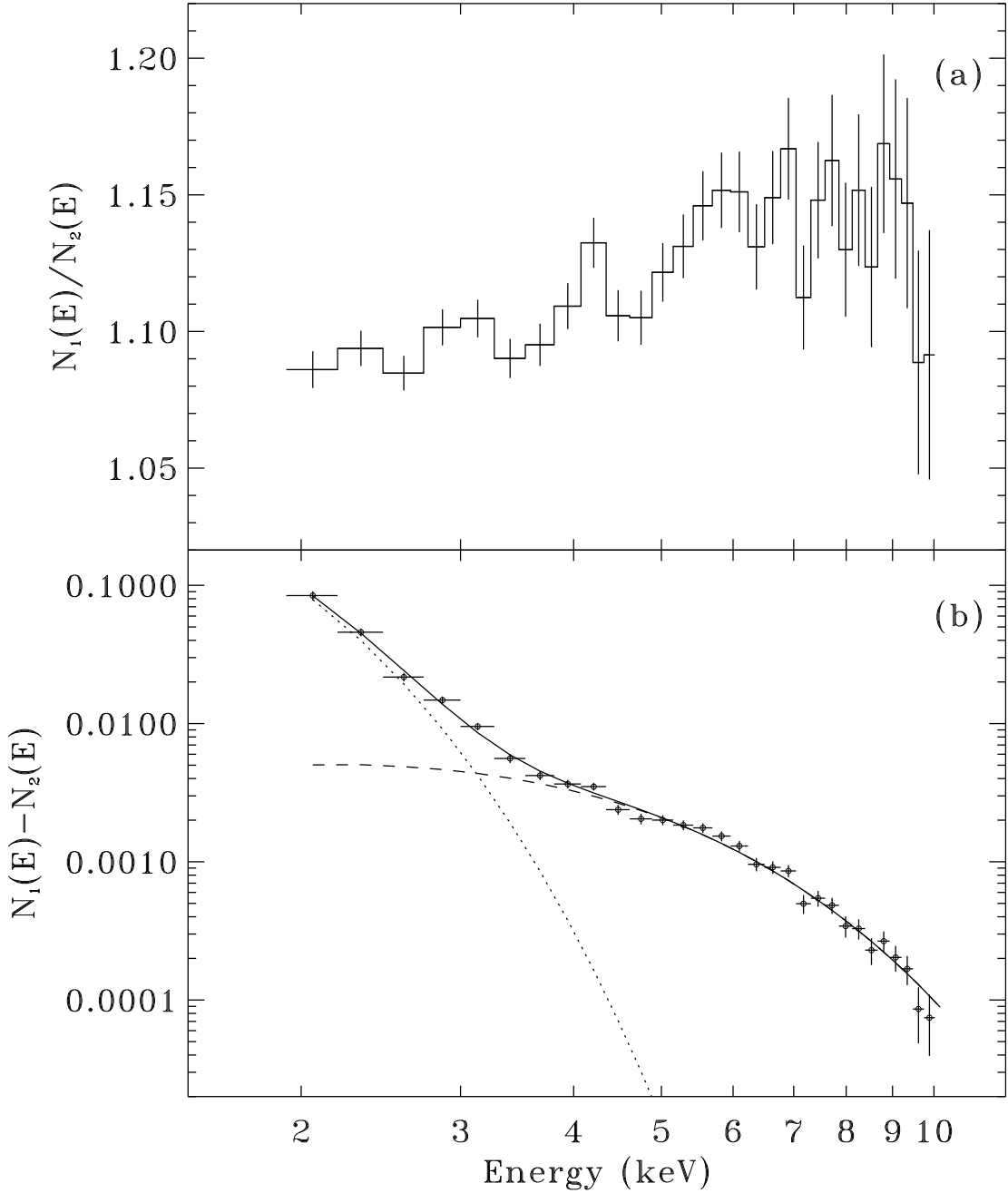}
\caption{Comparison of the energy spectra before the X-ray burst, 
$N_{1}(E)$ and after the X-ray burst, $N_{2}(E)$ in the energy 
range 2-10 keV. 
(a) Ratio of the two spectra ($N_{1}(E)$/$N_{2}(E)$); 
(b) Difference between the two spectra ($N_{1}(E)-N_{2}(E)$) in 
units of photons cm$^{-2}$ keV$^{-1}$ s$^{-1}$ (see Table 2). 
The model (solid line), the 0.28 keV component (dotted line) and 
the 1.15 keV component (dashed line) are also shown. }
\end{figure*}

\newpage
\begin{table} 
\scriptsize
\caption{Parameters of the spectra $N_{1}(E)$ and $N_{2}(E)$ 
(3-20 keV)\vspace{0.3cm}}
\begin{tabular}{|c|c|c|c|c|c|c|c|c|}
\hline
Spectrum &Model &$T^{(1)}_{1}$,  $T^{(2)}_{MCD}$(keV)
&$R^{(3)}_{1}$, $N^{(4)}_{MCD}$ & $T^{(1)}_{2}$(keV) & $R^{(3)}_{2}$ &Index$^{(5)}$ &$N^{(6)}_{PL}$ & ${\chi}^{2}$/DOF \\ \hline

$N_{1}(E)$	&2BB+PL&0.26$\pm$0.01	&274$\pm$47	&1.06$\pm$0.01&2.09$\pm$0.05&3.11$\pm$0.13	&0.37$\pm$0.22	&102.1/45 \\ \cline{2-9}
	&BB+MCD+PL&0.28$\pm$0.01 &(9.0$\pm$3.8)${10}^{4}$ &1.06$\pm$0.01&2.09$\pm$0.05 &3.08$\pm$0.13 &0.32$\pm$0.20 &107.3/45 \\ \hline
$N_{2}(E)$ 	&2BB+PL&0.25$\pm$0.01&326$\pm$155&1.02$\pm$0.01&2.14$\pm$0.07&2.65$\pm$0.14&0.15$\pm$0.15&61.9/45
\\ \cline{2-9}	
	&BB+MCD+PL&0.27$\pm$0.02&(1.3$\pm$0.9)${10}^{5}$  &1.02$\pm$0.01&2.13$\pm$0.07 &2.65$\pm$0.15 &0.16$\pm$0.16 &62.5/45 \\ \hline
\end{tabular}
\small

{\it Notes:} (1) BB temperature; (2) The inner disk temperature in MCD; 
(3) Apparent BB radii in km, assuming the distance to the source is 2.5 kpc; 
(4) The normalization of disk BB in MCD: 
(($R_{in}/km$)/(D/2.5kpc))$^{2}cos(\theta)$, where $R_{in}$ is the inner 
disk radius, D the distance to the source, and $\theta$ the angle of the disk;
(5) Photon spectral index; (6) photons cm$^{-2}$ keV$^{-1}$ s$^{-1}$ at 1 keV;
\end{table}

\begin{table}
\small
\caption{Model Fit to the subtracted spectrum $N_{1}(E)-N_{2}(E)$ (2-10 keV)}
\begin{tabular}{|c|c|c|c|c|c|c|}
\hline
$T_{1}$ (keV)&${R}^{(1)}_{1}$ (km)&
Flux1$^{(2)}$ &
$T_{2}$ (keV)&${R}^{(1)}_{2}$ (km)&
Flux2$^{(2)}$ &
${\chi}^{2}$/DOF  \\ \hline
0.28$\pm$0.03 & 42$\pm$12 & (3.5$\pm$1.4)$\times$10$^{-2}$ &
1.15$\pm$0.03 & 0.62$\pm$0.04 & (1.5$\pm$0.2)$\times$10$^{-2}$ &
31.0/25 \\ \hline
\end{tabular}
\small

{\it Notes: }
(1) Apparent BB radii in km for a distance of 2.5 kpc; 
(2) photons cm$^{-2}$ s$^{-1}$.
\end{table}
\end{document}